\documentclass[linenum]{SSA-BSSA}
\usepackage[utf8]{inputenc}
\usepackage{caption}
\usepackage{soul}
\usepackage{svg}
\usepackage{siunitx}
\citethisauthor{Anjali C. Dhabu, Aida Hejazi Nooghabi, and C\'{e}line Hadziioannou}
\doi{00.0000/0000000000}
\recdate{00 Month 0000}

\graphicspath{
	{./Figures/}
	{./logo/}
}
\nolinenumbers
\begin{document}
	
\title{Spectral Element Extension for Direct Simulation of Rotational Ground Motions}

\author[*1\orc{0000-0002-2913-3013}]{Anjali C. Dhabu}%
\author[1\orc{0009-0003-6580-3365}]{Aida Hejazi Nooghabi}
\author[1\orc{0000-0002-5312-2226}]{C\'{e}line Hadziioannou}

\affil[1]{Department of Earth System Sciences,Geophysics, University of Hamburg, Germany}{\auorc{0000-0002-2913-3013}{(ACD)}\auorc{0000-0002-5312-2226}{(CH)}}
\corau{*Corresponding author: anjali.dhabu@uni-hamburg.de}

\begin{abstract}
A complete characterization of earthquake induced ground motions require three translational and three rotational components, together with the six components of the strain tensor. Despite the importance of rotational ground motions, their characteristics are not yet completely recognized. Limited observational database and challenges of simulating them in realistic geological media has primarily constrained the research in this domain. In the present study, we develop an extension of the spectral-element package SPECFEM3D that directly evaluates rotational ground motion from spatial gradients of displacements at the Gauss-Lobatto-Legendre (GLL) integration points during the numerical solution. The implementation is validated against analytical solutions for a homogeneous elastic medium, showing good agreement in arrival times, amplitudes, waveform characteristics, and spatial propagation. For analyzing the propagation patterns of rotational ground motions, we introduce Peak Ground Rotation Rate (PGRR) maps, analogous to Peak Ground Velocity (PGV) maps. Comparison of PGRR and PGV maps revealed that the critical locations of translations and rotations do not coincide. Rotational ground motions are subsequently simulated in a layered medium for two earthquake source mechanisms, demonstrating the variation in PGRR maps for different source characteristics. Finally, a three-dimensional (3D) model of the Indo-Gangetic Basin is developed to show that low-velocity sedimentary deposits substantially amplify and redistribute rotational ground motions. The spatial distribution of PGRRs in the basin changes significantly relative to the corresponding generic layered medium. The developed framework provides a basis for investigating rotational ground motions in complex geological environments and assessing its potential relevance to sensor deployment strategies, seismic hazard and earthquake engineering.
\end{abstract}

\maketitle
\begin{keypoints}
	\item We extend SPECFEM3D to compute rotational motions directly and validate them against analytical solutions.
	\item We introduce Peak Ground Rotation Rate (PGRR) maps, the rotational analogue of peak ground velocity maps.
	\item PGRR maps vary with source mechanism and are strongly modified by low velocity sedimentary basin.
\end{keypoints}

\section{Introduction}
Modern seismology and earthquake engineering have traditionally relied on translational ground motions to characterize earthquake shaking and its effects on infrastructure \citep{aki2002quantitative, chopra2007dynamics, beroza1995seismic, Bormann2012, Tromp2019, Tkalcic2022}. These motions form the basis of Peak Ground Displacement (PGD), velocity (PGV), and acceleration (PGA) estimates which are fundamental to hazard assessment \citep{Nath2012, Gruenthal2018}. However, a complete description of earthquake-induced ground motion requires 12 degrees of freedom (DOF): three translational, three rotational and six strain components \citep{Agnew2014,Schmelzbach2018}. Although rotational components have long been recognized theoretically as an integral part of the seismic wavefield \citep{bouchon1982strain}, their composition, propagation, and engineering significance have received limited attention compared to translational motions. It was primarily the limited instrumentation technology that hindered their direct measurement \citep{Takeo1998, nigbor1994six}. Recent advances in the development of highly sensitive ring-laser gyroscopes \citep{Stedman1995,Pancha2000, schreiber2006geosensor} and fibre-optic rotational sensors \citep{Lee2009, Bernauer2012, Bernauer2018} have enabled direct observations of earthquake-induced rotational motions. Even then, continuous observations of rotations remain sparse with only a few permanent observatories worldwide which provide continuous rotational recordings \citep{Salvermoser2017}. The sparse availability of portable rotational sensors also hinders the acquisition of dense near-field rotational datasets.

Recent studies have demonstrated that rotational observations can provide constraints on seismic-wave type and polarization \citep{Sollberger2020}, improve earthquake source characterization \citep{Chen2023, Xu2025}, and reveal aspects of wavefield complexity that cannot be fully resolved from translational recordings alone. The implications of observed rotations also extend to earthquake engineering. Earthquake damage surveys frequently report substantial torsional response and structural failures, which are conventionally attributed to structural asymmetry, irregularity, or construction deficiencies \citep{newmark1969torsion, Chandler1986}. With increasing recognition of rotational ground motions generated by earthquakes, the rotational component of the incoming seismic wavefield may also significantly contribute to the torsional response of structures. This effect is expected to be particularly relevant in the near field, where strong spatial gradients in the seismic wavefield can produce appreciable rotational motions.

Current hazard assessment and seismic design practices overlook the contribution of rotations due to lack of observed near-field rotational ground motions. In the absence of recorded data, analytical approaches have been used to investigate near-field rotational ground motions within different continuum formulations, including (i) classical elastic media \citep{Cao2019, Singla2019, Cao2024}, (ii) reduced micropolar media \citep{Dhabu2019, Dhabu2021a, Dhabu2021}, and (iii) micropolar media \citep{Gade2015}. These studies are generally restricted to idealized Earth models, such as homogeneous or horizontally layered media, and cannot readily incorporate realistic three-dimensional (3D) geological structures, including sedimentary basins, lateral heterogeneity, and topographic effects. Numerical wave-propagation methods overcome these limitations and permit simulations in arbitrarily complex geological environments. However, simulation of rotations using existing numerical approaches requires computation of finite-difference approximations of translations between neighboring receiver locations \citep{Smerzini2009}. Such array-derived rotations (ADR) depend strongly on receiver spacing and array geometry and they often differ from point rotational measurements \citep{Suryanto2006, Lin2012, Gueguen2020, Taylor2021}. It is particularly relevant within the near field, where displacement gradients vary rapidly. As a result, the existing frameworks fail to provide reliable point rotational ground motion in realistic 3D media. Without such a framework, it is neither possible to quantify how rotations behave in complex geology nor assess whether they warrant inclusion in hazard assessment and design.
 
These limitations give rise to the following questions: (i) How can rotational ground motions be computed directly and accurately within a 3D numerical wave-propagation framework? (ii) Do rotational ground motions exhibit different propagation characteristics from those of conventional translational motions? (iii) How do rotational motions behave in the near field and how does their propagation inform earthquake engineering? and (iv) How are these characteristics influenced by earthquake source mechanism and realistic geological structures such as sedimentary basins? The present study addresses these questions by extending the open-source spectral-element package SPECFEM3D \citep{Komatitsch2024} to compute rotational ground motions directly from the displacement gradients at the Gauss--Lobatto--Legendre (GLL) integration points during the numerical solution. The implementation thereby provides point rotations together with displacements, velocities, and accelerations. To ensure the robustness of this development, it is first validated against analytical solutions for a homogeneous elastic medium. In the next step, the validated framework is employed to investigate the wavefield composition, propagation characteristics and response of the medium to rotational ground motions. 

Building upon these observations, the concept of Peak Ground Rotation Rate (PGRR) maps is introduced as the rotational analogue of conventional PGV maps to characterize the spatial distribution of peak rotational demand. The influence of earthquake source mechanism on PGRR is investigated for homogeneous and layered media. Finally, the methodology is applied to a realistic 3D model of the Indo-Gangetic Basin to examine the amplification and propagation of rotational ground motions in a complex sedimentary environment.

\section{Implementation of Rotational Motions in SPECFEM3D}
Among the available open-source computational tools for seismic wave propagation, SPECFEM3D \citep{Komatitsch2024}  is one of the most widely used software packages. It employs the spectral-element method \citep{Komatitsch2002, Komatitsch2002a}, which combines the geometric flexibility of finite elements with the high accuracy of spectral techniques. The code computes seismic wavefields by solving the elastodynamic equations on a mesh discretized using GLL points. Ground displacements and their corresponding time derivatives (velocities and accelerations) are computed at each GLL point and recorded at user-defined receiver locations through interpolation. Within each spectral element, the displacement field is represented using the GLL interpolation functions as:
\begin{equation}
	\vec{{u}}({x},t) = \sum_{a=1}^{N_{\mathrm{GLL}}}\phi_a(\xi,\eta,\zeta)\vec{{u}}_a(t),
	\label{eq:sem_interp}
\end{equation}
where $N_{\mathrm{GLL}}$ is the total number of GLL interpolation points in an element, $\phi_a$ denotes the Lagrange interpolation function associated with the $a^{\mathrm{th}}$ GLL point oriented in the direction ($\xi, \eta ,\zeta $) and $\vec{{u}}_a (t)$ represents the displacement vector at that point at time instant $t$. Translational ground motions are subsequently obtained at user-defined receiver locations through spectral-element interpolation of the corresponding wavefields.

Rotational ground motions are related to the displacement field through the curl operator:
\begin{equation}
	\vec{\Theta}=\frac{1}{2}\nabla\times\vec{u}
	\label{rot_basic_eqn}
\end{equation}  
where $\vec\Theta$ is the rotational wavefield represented in terms of three orthogonal components as $\vec{\Theta}=\Theta_x \hat{i}+\Theta_y \hat{j}+\Theta_z \hat{k}$ and $\vec{u}$ is the displacement wavefield represented as $\vec{u}=u_x \hat{i}+u_y \hat{j}+u_z \hat{k}$. In both the vector representations, $\hat{i}, \hat{j} \text{ and } \hat{k}$ are the unit vectors in the direction of $x-$, $y-$ and $z-$ axis, respectively. Substituting the vector form of $\vec{u}$ in eq.\ref{rot_basic_eqn}, the following equation for $\vec{\Theta}$ is obtained:
\begin{equation}
	\vec{\Theta}=\frac{1}{2}\left(\left(\frac{\partial u_z}{\partial y}- \frac{\partial u_y}{\partial z}\right)\hat{i}+\left(\frac{\partial u_x}{\partial z}- \frac{\partial u_z}{\partial x}\right)\hat{j}+\left(\frac{\partial u_y}{\partial x}- \frac{\partial u_x}{\partial y}\right)\hat{k}\right)
	\label{rot_full_eqn}
\end{equation}  
%where
%\begin{subequations}
%	\begin{equation}
%		\Theta_x = \frac{1}{2}\left(\frac{\partial u_z}{\partial y}- \frac{\partial u_y}{\partial z}\right)
%	\end{equation}
%	\begin{equation}
%		\Theta_y = \frac{1}{2}\left(\frac{\partial u_x}{\partial z}- \frac{\partial u_z}{\partial x}\right)
%	\end{equation}
%	\begin{equation}
%		\Theta_z = \frac{1}{2}\left(\frac{\partial u_y}{\partial x}- \frac{\partial 	u_x}{\partial y}\right)
%	\end{equation}
%	\label{eq:rotation components}
%\end{subequations}
%For example, rotation about the $x$-axis, $\Theta_x$, depends on the spatial gradients of the $y$- and $z$-components of displacement, whereas it is independent of the displacement component $u_x$ itself. Similarly, $\Theta_y$ is determined by the spatial gradients of $u_x$ and $u_z$, while $\Theta_z$ depends on those of $u_x$ and $u_y$.
Thus, each rotational component is governed by the spatial variation of the two displacement components perpendicular to its axis of rotation.  

%Rotational ground motions therefore quantify the local differential deformation and twisting of the displacement wavefield rather than representing particle motion along a particular direction. 

In the present work, SPECFEM3D is extended to compute rotational ground motion directly from the spatial gradients of the numerically simulated displacement field. Differentiating equation (\ref{eq:sem_interp}) with respect to the physical coordinates provides:
\begin{equation}
	\nabla\vec{{u}}=\sum_{a=1}^{N_{\mathrm{GLL}}}\left(\nabla\phi_a\right) \otimes\vec{{u}}_a,
	\label{eq:grad_u}
\end{equation}

where $\otimes$ denotes the dyadic product. Substituting equation (\ref{eq:grad_u}) into equation (\ref{rot_basic_eqn}) yields:
\begin{equation}
	\vec{\Theta}=\frac{1}{2}\sum_{a=1}^{N_{\mathrm{GLL}}}\left(\nabla\phi_a\right)
	\times\vec{{u}}_a.
	\label{eq:sem_rotation}
\end{equation}
Equation (\ref{eq:sem_rotation}) provides a direct evaluation of the rotational field from the spatial derivatives of the spectral-element interpolation functions and the displacement field at the GLL points. The rotational field is evaluated at the GLL points at every time step, immediately after the displacement field is computed, and is interpolated to the user-defined receiver locations using the same spectral-element interpolation that is employed for translational quantities. Rotations about the East-West, North-South, and Vertical axes are then obtained directly alongside displacements, velocities, and accelerations without requiring external post-processing. This implementation provides a physically consistent, mesh-based evaluation of rotational ground motions that avoids the receiver-spacing dependence inherent to finite-difference post-processing approaches.
	
\subsection{Validation of the Numerical Implementation}
\begin{figure} % fig1
	\centering
	\includegraphics[width=0.5\textwidth, keepaspectratio]{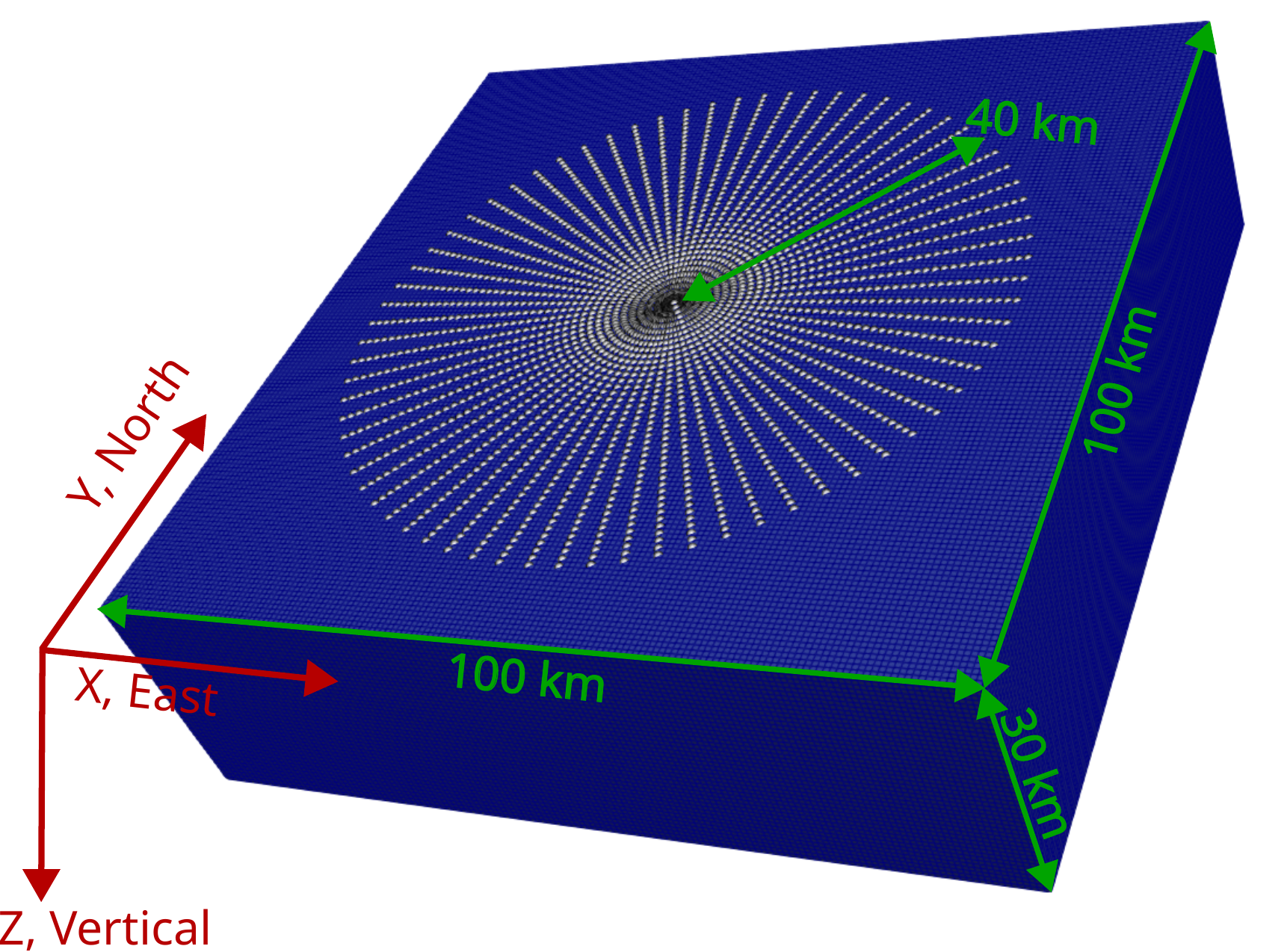}
	\caption{Numerical model and source reciever set-up for a homogeneous medium. Uniform colour signifies homogeneous medium and the top concentric circles signify the receiver locations.}
	\label{Homogeneous_mesh}
\end{figure}

\begin{figure*} % fig1
	\global\twocolcaptionfalse
	\centering
	\includegraphics[width=0.8\textwidth, keepaspectratio]{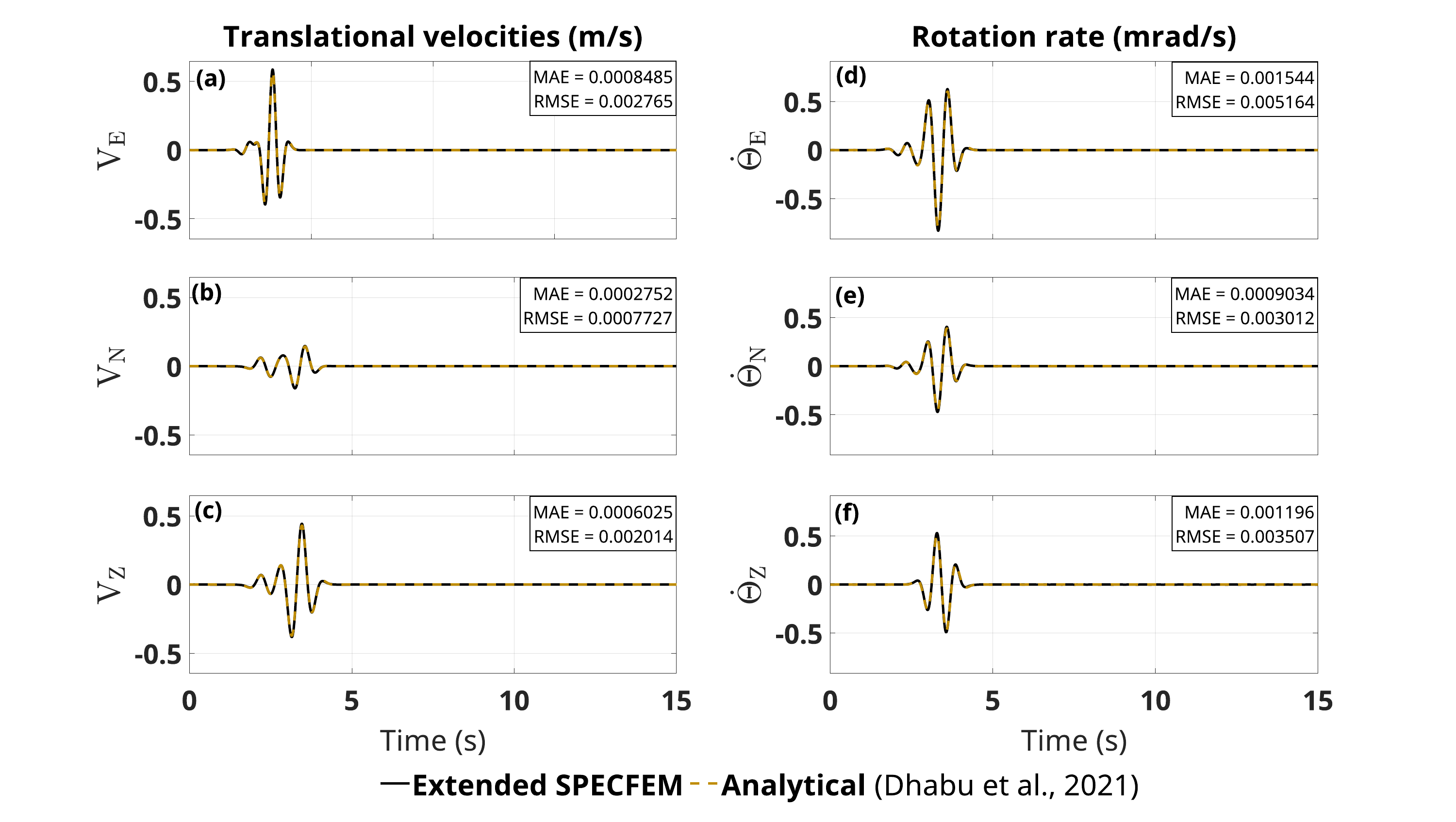}
	\caption{Comparison of \textbf{(a--c)} translational velocities (in m/s) and \textbf{(d--f)} rotational rate (in mrad/s) about the East, North and Vertical direction respectively for a station located at an epicentral distance of \SI{5}{\km} and \SI{120}{\degree} obtained from SPECFEM3D (black line) and \citet{Dhabu2021} (yellow dashed line).}
	\label{one_reciever_simulations}
\end{figure*}

The simulated rotational motions from the developed version of SPECFEM3D are validated against available analytical solutions to verify the correctness of the implementation. A rectangular domain measuring \SI{100}{\km} $\times$ \SI{100}{\km} in the horizontal directions and extending to a depth of \SI{30}{\km} is modeled as a homogeneous half-space (Figure \ref{Homogeneous_mesh}).
The medium is characterized by a P-wave velocity ($V_p$) of \SI{5600}{\m/\s}, an S-wave velocity ($V_s$) of \SI{3200}{\m/\s}, and a density ($\rho$) of \SI{2600}{\kg/\m^3}. The domain consists of $1,143,792$ spectral elements and is discretized to resolve periods down to \SI{0.25}{\s}. The half-space is subjected to a pure strike-slip seismic source with a strike angle of \SI{0}{\degree}, dip angle of \SI{90}{\degree} and rake angle of \SI{0}{\degree}. The source is modeled as a double-couple, with a seismic moment of $10^{24}$ dyne-cm and a Ricker source-time function having a central frequency of \SI{1}{\Hz}. This source configuration is expected to have a maximum frequency of \SI{2.5}{\Hz} which can be conveniently simulated by the developed model. The source is located at the center of the domain at a depth of \SI{5}{\km}. Receivers are distributed along concentric circles on the free surface, centered above the source location, with radial and azimuthal spacings of \SI{1}{\km} and \SI{5}{\degree}, respectively. The receiver distribution extends to an epicentral distance of \SI{40}{\km}. This configuration enables the assessment of the spatial decay and radiation characteristics of rotational ground motions. Receivers beyond this distance are excluded to minimize the influence of artificial reflections arising from the computational boundaries. Figure (\ref{Homogeneous_mesh}) illustrates the homogeneous spectral element model and the corresponding source–receiver geometry. The simulated ground motions are filtered within the frequency range of \SI{0.25}{\Hz} to \SI{2.5}{\Hz}, consistent with the frequency content of the source-time function. Since rotational seismometers typically measure rotation rates rather than rotations, the rotation vector components obtained from SPECFEM3D implementation are differentiated with respect to time to obtain rotational rates. Figure (\ref{one_reciever_simulations}) presents the translational velocities and rotational rates simulated using the developed SPECFEM3D implementation for a receiver located at an epicentral distance of \SI{5}{\km} and an azimuth of \SI{120}{\degree}.

\begin{figure*} % fig1
	\global\twocolcaptionfalse
	\centering
	\includegraphics[width=\linewidth, keepaspectratio]{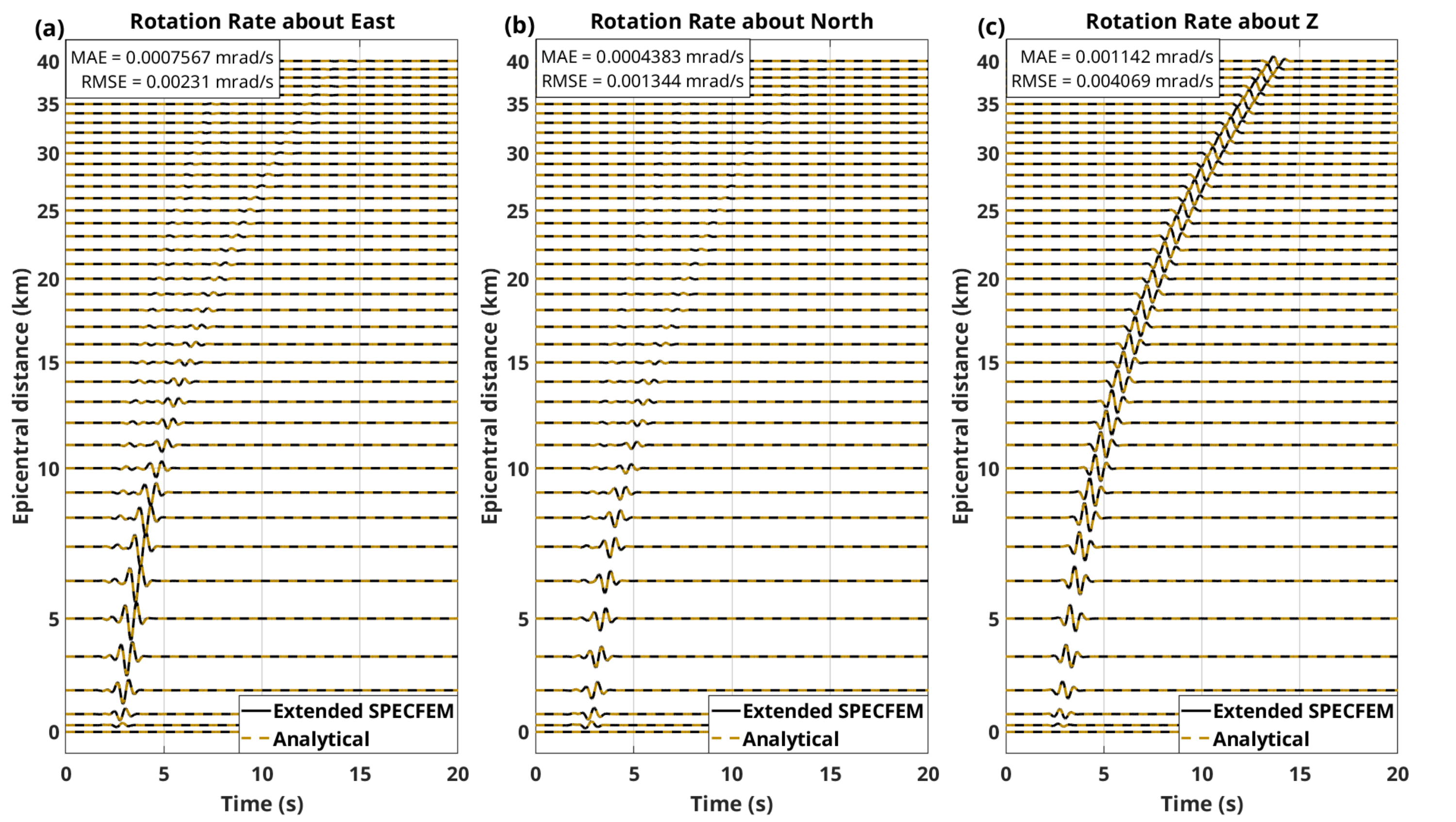}
	\caption{Comparison of rotational rate simulated about the \textbf{(a)} East \textbf{(b)} North, and \textbf{(c)} Vertical axes for 41 stations placed at an interval of \SI{1}{\km} each along an azimuthal direction of \SI{60}{\degree}. }
	%\caption{Test caption.}
	%\caption{Comparison of rotational rate simulated about the East, North and Vertical axis for 20 stations placed at an interval of \SI{1}{\km} each along an azimuthal direction of 60$\circ$.}
	\label{multiple_reciever_simulations}
\end{figure*}

The simulated ground motions are validated against the analytical formulation developed by \citet{Dhabu2021}. They developed analytical solutions for seismic-wave propagation were derived under different physical assumptions and implemented numerically to compute translational velocities and rotational rates. Although the underlying formulations in \citet{Dhabu2021} differ from the classical elastic framework considered in the present study, the analytical model can be configured to represent the same homogeneous elastic medium by selecting appropriate material parameters. This enables a direct comparison between the analytical model of \citet{Dhabu2021} and numerical solutions obtained from the current implementation. Translational velocities and rotational rates are simulated up to a frequency of \SI{2.5}{\Hz} from the analytical model for the same medium and source characteristics as considered in the numerical model and compared with the results obtained from the extended SPECFEM3D implementation. Figure (\ref{one_reciever_simulations}) presents the comparison of the three components of translational velocity (Figure \ref{one_reciever_simulations}a--\ref{one_reciever_simulations}c) and rotational rate (Figure \ref{one_reciever_simulations}d--\ref{one_reciever_simulations}f) for the same receiver. The numerical and analytical solutions exhibit excellent agreement in terms of the arrival times, waveform characteristics, and dominant frequency content. The observed mean absolute error (MAE) and root mean square error (RMSE) for each of the waveform is reported in Figure (\ref{one_reciever_simulations}a--\ref{one_reciever_simulations}f). This demonstrates that the implemented formulation correctly captures the rotational component of the propagating seismic wavefield. The simulated amplitudes also show good agreement with the analytical solutions. Ground motions are also compared at all the receiver locations. For the purpose of brevity, comparison of rotation rate about horizontal axes (hereafter referred as tilt motions) and about the vertical axis (hereafter referred as torsional motion) for receivers up to an epicentral distance of \SI{40}{\km} along the azimuthal direction of \SI{60}{\degree} is shown in Figure \ref{multiple_reciever_simulations}.
The MAE for rotational motions over these 41 stations about East, North and Vertical direction is observed to be \SI{7.5674e-4}{mrad/s}, \SI{4.38e-4}{mrad/s}, and \SI{1.1424e-3}{mrad/s} respectively. The RMSE is also reported in the respective figures. The figure demonstrates that the agreement between the analytical and numerical solutions is maintained over a broad range of epicentral distances.

The agreement between the analytical and numerical solutions demonstrates that the extension accurately reproduces the rotational characteristics of seismic waves propagating in a homogeneous elastic medium. The implementation reproduces the expected spatial decay, geometric spreading, and waveform evolution with increasing source–receiver distance. The consistency of the amplitudes of rotational-rate and arrival characteristics across all receiver locations indicates that the implemented formulation accurately computes rotational motions throughout the simulated wavefield and is not limited to a specific source-receiver configuration. This validation confirms the accuracy of the rotational-motion calculation within SPECFEM3D and supports its application to more complex 3D simulations. The ability to compute rotational ground motions directly within the spectral-element solver enables their simulation in realistic 3D geological models where analytical solutions are no longer applicable. This capability now provides a foundation for investigating earthquake source characterization, basin amplification, soil–structure interaction, and structural health monitoring using both translations and rotations.

\section{Rotations, their composition and propagation}
The successful validation of the numerical implementation enables a systematic investigation of the physical characteristics of rotational ground motions. Unlike translational measurements, rotational motions are governed by the spatial gradients of the seismic wavefield and therefore contain additional information regarding wave propagation, source radiation, and deformation. The present section showcases the capabilities of new SPECFEM3D by first examining the wavefield composition of the rotational components and subsequently investigating their dependence on earthquake source mechanism and propagation patterns.

\begin{figure*} % fig1
	\global\twocolcaptionfalse
	\centering
	\includegraphics[width=0.75\textwidth]{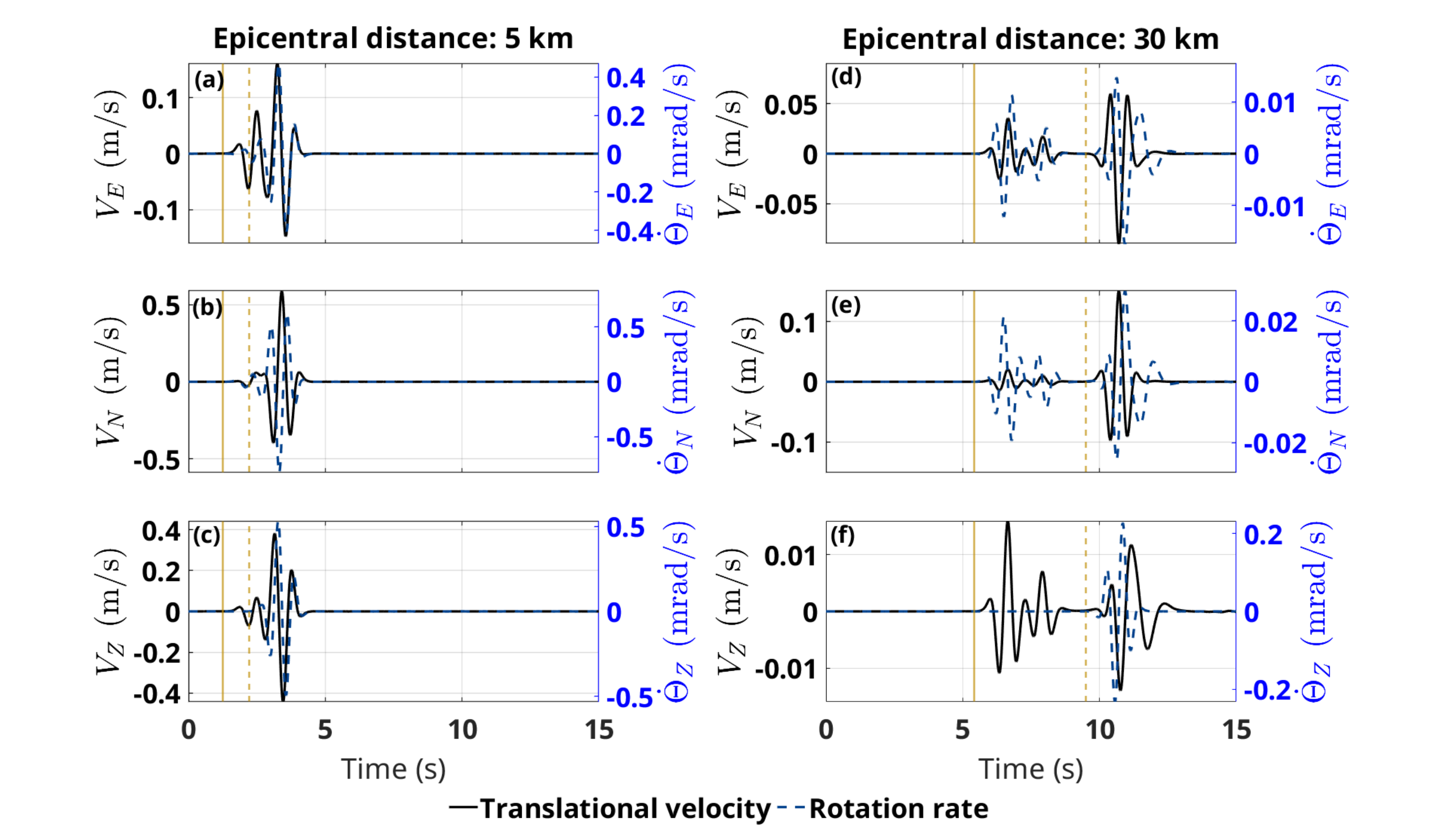}
	\caption{Comparison of the three components of translational velocity (black) and rotational-rate waveforms (blue) at epicentral distances of \textbf{(a--c)} \SI{5}{\km} and \textbf{(d--f)} \SI{35}{\km}, illustrating a clear separation of seismic wave types with increasing propagation distance. The solid and dashed yellow lines in each panel indicate the approximate arrival times of P and S- waves respectively.}
	\label{wavetype_comparison}
\end{figure*}

To examine the wavefield composition, translational velocity and rotational-rate waveforms are compared at epicentral distances of \SI{5}{\km} (Figure \ref{wavetype_comparison}a--\ref{wavetype_comparison}c) and \SI{35}{\km} (Figure \ref{wavetype_comparison}d--\ref{wavetype_comparison}f) along an azimuth of \SI{60}{\degree}. 
At \SI{5}{\km}, the P-, SV-, and SH-wave arrivals overlap considerably, so individual phases cannot be separated in both translational and rotational records. At \SI{35}{\km}, geometric spreading separates the wave packets. The torsional rotation rate is dominated by the SH-wave (Figure \ref{wavetype_comparison}f), whereas the tilt rotation rates consist of both P- and SV-wave contributions (Figure \ref{wavetype_comparison}d--\ref{wavetype_comparison}e). These observations are consistent with the analytical findings \citep{Singla2019}, confirming that torsional motion isolates horizontally polarized shear-wave energy without additional wavefield decomposition, whereas the tilt components remain sensitive to P- and SV- waveforms.

These observations highlight that rotational measurements offer considerable potential for wavefield decomposition, automated phase identification, and improved earthquake source characterization. The seismic hazard assessment is however primarily concerned with the peak amplitudes of ground motion and their spatial distribution. It is therefore important to investigate how peak rotational motions propagate from the source and how the source mechanism influences the spatial distribution of rotations.
	
\subsection{Propagation patterns of Rotational Ground Motions}
\begin{figure*} % fig1
	\global\twocolcaptionfalse
	\centering
	\includegraphics[width=\textwidth]{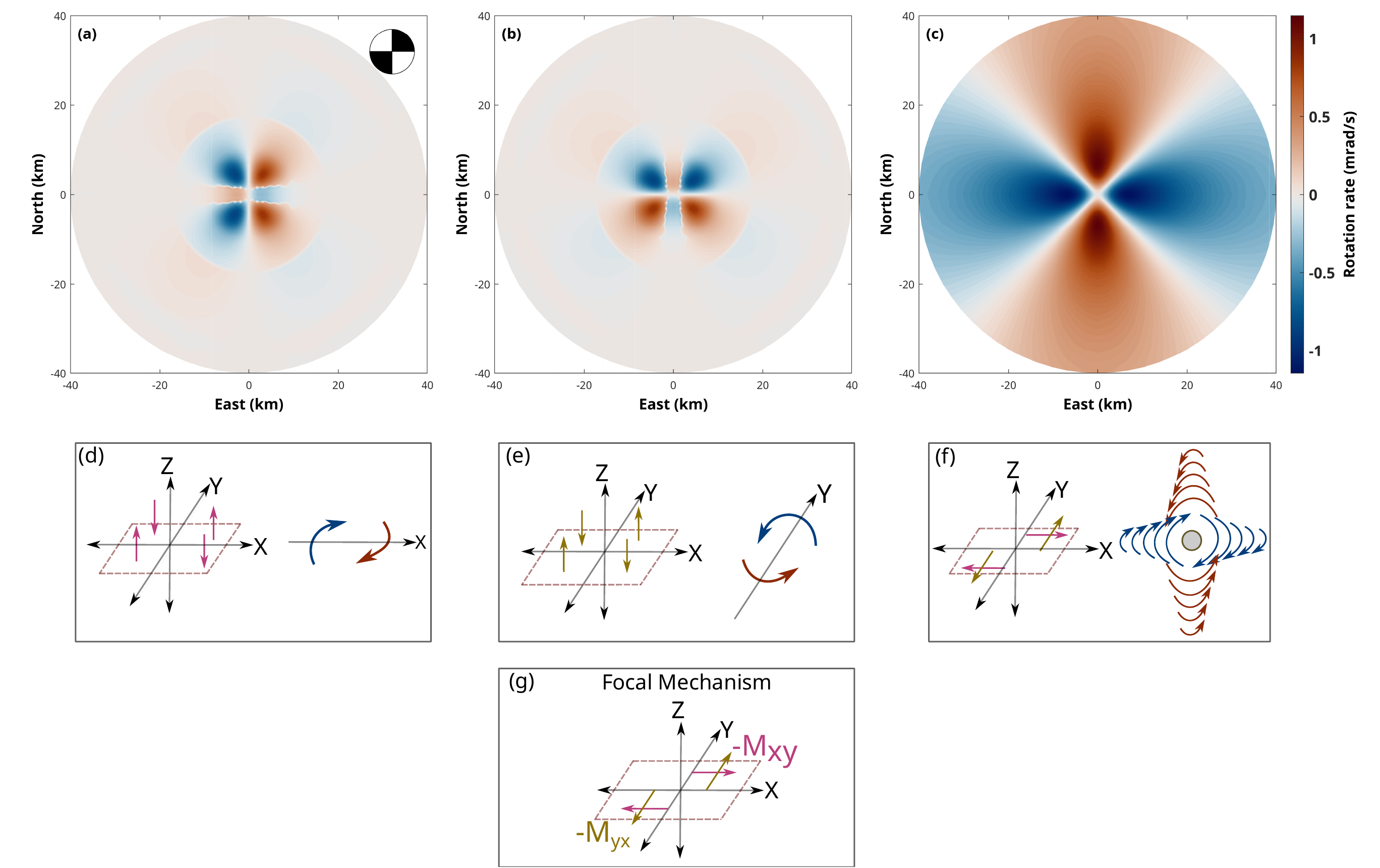}
	\caption{Contour maps of the peak amplitude of rotational ground motion about the East \textbf{(a)} , North \textbf{(b)} and Vertical Axis \textbf{(c)}. \textbf{(d--f)} Schematic illustration of the displacement gradients and force couples responsible for generating rotations about the East, North, and Vertical axes respectively for a strike-slip source. \textbf{(g)} Focal mechanism of the strike-slip earthquake.}
	\label{abs_signed_comparison}
\end{figure*}

In the present study, the propagation characteristics of rotational ground motions about the three orthogonal directions are investigated. Figure (\ref{abs_signed_comparison}) presents the peak tilt and torsional amplitudes obtained at each receiver location for the homogeneous model.
The simulation set-up and the source parameters are same as those chosen for validating the rotational ground motions in the previous section. The maximum amplitude of the strong ground motion along with the positive or negative sign of the rotation rate waveform is identified as the peak rotation rate. The contour plots (Figure \ref{abs_signed_comparison}a--\ref{abs_signed_comparison}c) preserve the sense of rotation (clockwise or counter-clockwise), thereby revealing the nodal structure of the rotational wavefield and the symmetry associated with the strike-slip double-couple source. Consequently, they retain physical information regarding source radiation and deformation. The physical origin of these rotational patterns is illustrated schematically in Figures (\ref{abs_signed_comparison}d)--(\ref{abs_signed_comparison}g). 

A left-lateral strike-slip earthquake is represented by two orthogonal double couples, $-M_{xy}$ and $-M_{yx}$ (Figure \ref{abs_signed_comparison}g), each producing characteristic displacement gradients that generate rotations about the three coordinate axes. Considering the contribution of the $-M_{xy}$ double couple, the imposed horizontal shear causes compression in two diagonally opposite quadrants and extension in the remaining two quadrants. The resulting vertical displacement field, represented in the left panel of Figure (\ref{abs_signed_comparison}d), consists of upward motion within the compressional quadrants and downward motion within the extensional quadrants. This establishes opposite vertical displacement gradients across the fault. These gradients generate positive rotations about the X-axis on the right-hand side of the nodal plane and negative rotations on the left-hand side (represented in the right panel of Figure \ref{abs_signed_comparison}d). This phenomenon produces the positive and negative lobes observed in the contour map shown in Figure (\ref{abs_signed_comparison}a). An analogous interpretation applies to the remaining rotational components, where the polarity and spatial distribution of the rotational motions directly reflect the displacement gradients generated by the two orthogonal double couples. 

These rotational contour maps that preserve the sign of peak rotational rate amplitude establish a direct physical link between the earthquake source mechanism, the resulting displacement gradients, and the spatial distribution of rotational ground motions. While these maps are invaluable for understanding the physical origin of rotational wavefields, seismic hazard assessment is generally concerned with the maximum rotational demand irrespective of its polarity.

\subsection{Peak Ground Rotation Rate Maps}
\begin{figure*} % fig1
	\global\twocolcaptionfalse
	\centering
	\includegraphics[width=\linewidth, keepaspectratio]{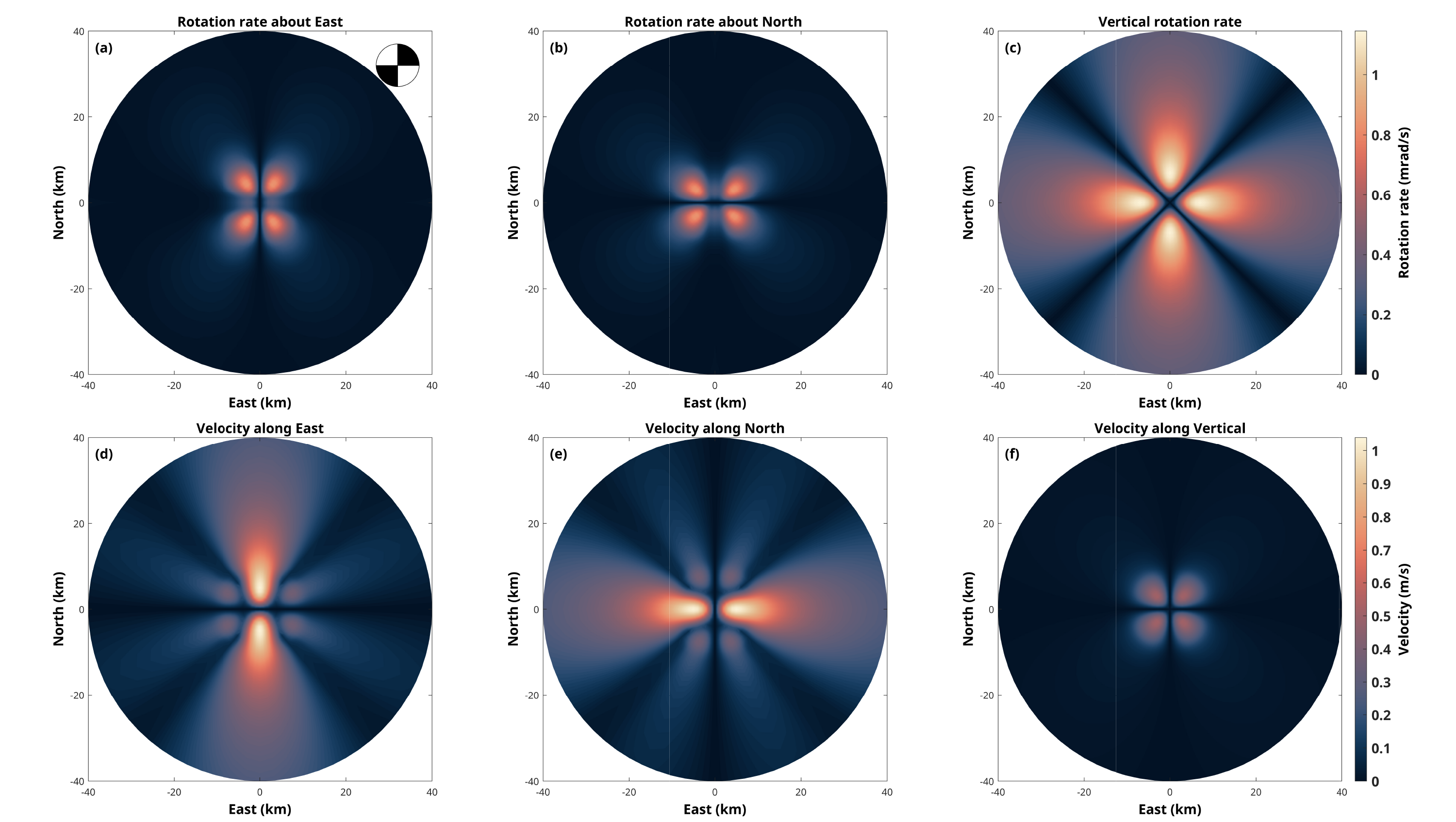}
	\caption{Comparison of the amplitudes and propagation pattern of PGRR and PGV maps obtained for the generic layered medium when subjected to a strike-slip earthquake of 10$^{24}$ dyne-cm seismic moment. \textbf{(a)--(c)} show PGRR maps about the East, North and vertical axis and \textbf{(d)--(f)} show PGV maps along the Earth, North and vertical axis.}
	\label{PGV_vs_PGRR}
\end{figure*}

For engineering applications, the amplitude and location of peak rotational ground motions plays a more crucial role in quantifying the rotational demand of structures. Accordingly, PGRR is defined herein as the maximum absolute value of the rotational-rate time history, contrary to the previous section which considers maximum value along with its sign. The resulting PGRR maps identify the maximum rotational demand at every receiver location and therefore represent the rotational analogue of conventional PGV or PGA maps. Figure (\ref{PGV_vs_PGRR}) shows the PGRR map for the same source mechanism as considered in the previous section and compares them with the conventional PGV maps. 	
Although PGV and PGRR are derived from the same propagating seismic wavefield, they exhibit markedly different spatial characteristics as they represent different physical quantities. The PGV maps (Figures \ref{PGV_vs_PGRR}d--\ref{PGV_vs_PGRR}f) describe the maximum translational particle velocity at each receiver and largely reflect the radiation pattern of the earthquake source. In contrast, the PGRR maps (Figures \ref{PGV_vs_PGRR}a--\ref{PGV_vs_PGRR}c) represent the maximum rotational response arising from the spatial gradients of the displacement wavefield. Thus they emphasize regions of rapid spatial deformation rather than particle motion at specific points.

A comparison of the propagation patterns reveals that the regions of maximum translational motion do not coincide with those experiencing the largest rotational motions. The horizontal PGV components (Figure \ref{PGV_vs_PGRR}d--\ref{PGV_vs_PGRR}e) exhibit predominantly double-lobed radiation patterns aligned with the nodal and auxiliary planes of the fault. In contrast, the corresponding tilt components (Figure \ref{PGV_vs_PGRR}a-\ref{PGV_vs_PGRR}b) display well-defined quadruple radiation pattern. The zero-rotation nodal lines separating adjacent lobes coincide with the nodal plane for rotation about the East axis (Figure \ref{PGV_vs_PGRR}a)  and with the auxiliary plane for rotation about the North axis (Figure \ref{PGV_vs_PGRR}b). The vertical components also exhibit fundamentally different behaviour. While both the vertical PGV (Figure \ref{PGV_vs_PGRR}f) and vertical PGRR (Figure \ref{PGV_vs_PGRR}c) exhibit quadruple-lobed radiation patterns, the maxima of the vertical PGRR are aligned with the nodal and auxiliary plane. On the contrary, the maxima of the vertical PGV occur in different azimuthal directions with zero PGV occurring along the nodal and auxiliary plane. These observations demonstrate that rotational motions encode different aspects of the source radiation and wavefield deformation than conventional translational measurements. The relative amplitudes of the translational and rotational components also differ significantly. For the source mechanism considered here, the largest translational motions occur along the horizontal directions, whereas the vertical translational velocities remain comparatively small. In contrast, the torsional component exhibits the largest PGRR amplitudes, indicating that torsional ground motion dominates over the tilt components. Since torsional motion can induce significant rotational demand in structures, this observation suggests that the component governing rotational seismic demand may differ from that governing translational seismic demand. Although the present analysis does not directly quantify structural damage, it demonstrates that rotational ground motions provide complementary information regarding seismic-wave propagation and deformation that cannot be inferred from translational measurements alone. Consequently, incorporating rotational ground motions into seismic hazard assessment has the potential to improve the characterization of earthquake loading, particularly for structures that are sensitive to torsional excitation. 
	
\section{Rotational Ground Motions in a Layered Medium}
The analysis presented for the homogeneous medium establishes the fundamental characteristics of rotational wave propagation and provides a physical interpretation of the observed PGRR radiation patterns. However, a homogeneous medium does not adequately represent the Earth's crust, where seismic waves undergo reflection, refraction, mode conversion, and scattering across interfaces. These processes modify the spatial gradients of the displacement wavefield and are therefore expected to influence rotational motions in a different way as compared to translational motions. The present section investigates the propagation characteristics of rotational ground motions when a layered Earth medium is subjected to different earthquake source mechanisms. 

\begin{figure} % fig1
	\centering
	\includegraphics[width=0.5\textwidth, keepaspectratio]{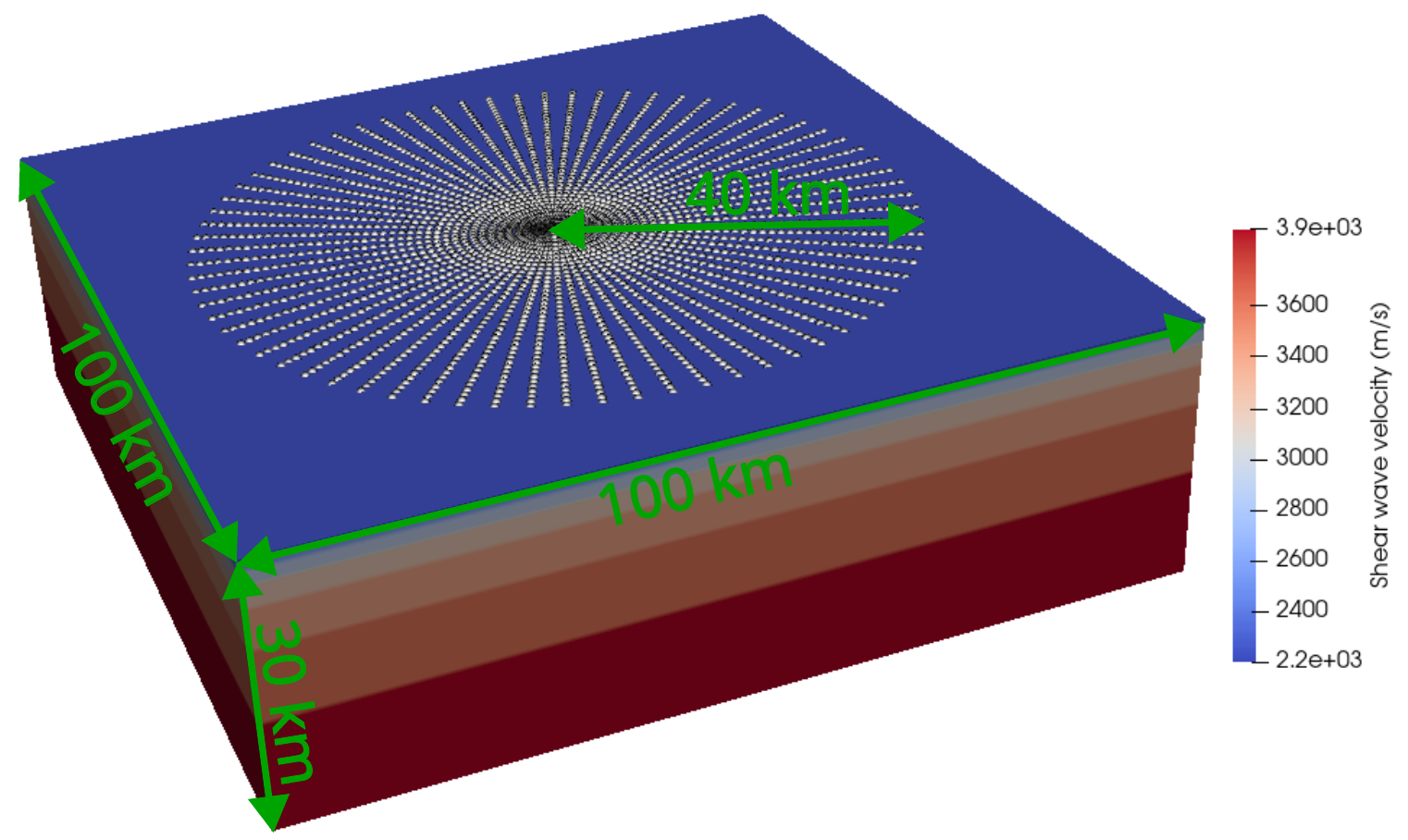}
	\caption{Numerical model and source receiver set-up for the layered medium. The colors signify the different S-wave velocities and the top concentric circles signify the receiver locations.}
	\label{Layered_mesh}
\end{figure}

A seven-layer elastic Earth model is considered for this purpose, with the material properties summarized in Table~\ref{layered_properties}. The numerical model employs the same source--receiver configuration as used for the homogeneous-medium simulations (Figure~\ref{Layered_mesh}) to facilitate a direct comparison between the homogeneous and layered models.
\begin{table}
	\tbl{Material properties of the layered Earth medium\label{layered_properties}}%
	{\begin{tabular*}{\columnwidth}{@{\extracolsep{\fill}}ccccc@{}} %
			Layer & Depth & $V_s$ & $V_p$ & Density    \\
			& (km)  & (m/s) & (m/s) & (kg/m$^3$) \\ \hline
			% \multicolumn{1}{l}{} & \textbf{12856} & \textbf{23.2 \%} & \textbf{0.2 \%} & \textbf{18.8 \%} \\
			\colrule
			1    & 0 -- 0.5     & 2200  & 4000  & 2200       \\
			2    & 0.5 -- 1.5   & 2600  & 4700  & 2350       \\
			3    & 1.5 -- 3.0   & 2900  & 5200  & 2500       \\
			4    & 3.0 -- 5.5   & 3200  & 5600  & 2600       \\
			5    & 5.5 -- 10.0  & 3400  & 5900  & 2650       \\
			6    & 10.0 -- 18.0 & 3600  & 6300  & 2750       \\
			7    & 18.0 -- 30.0 & 3800  & 6600  & 2850       \\
			\botrule
	\end{tabular*}}
\end{table}
The model consists of approximately $2.4$ million spectral elements and resolves seismic waves with periods up to \SI{0.25}{\second}. Displacements and rotations are simulated using the same numerical parameters as adopted for the homogeneous medium, and the resulting waveforms are band-pass filtered between \SI{0.25}{\hertz} and \SI{2.5}{\hertz}.

\begin{figure*} % fig1
	\global\twocolcaptionfalse
	\centering
	\includegraphics[width=\textwidth, keepaspectratio]{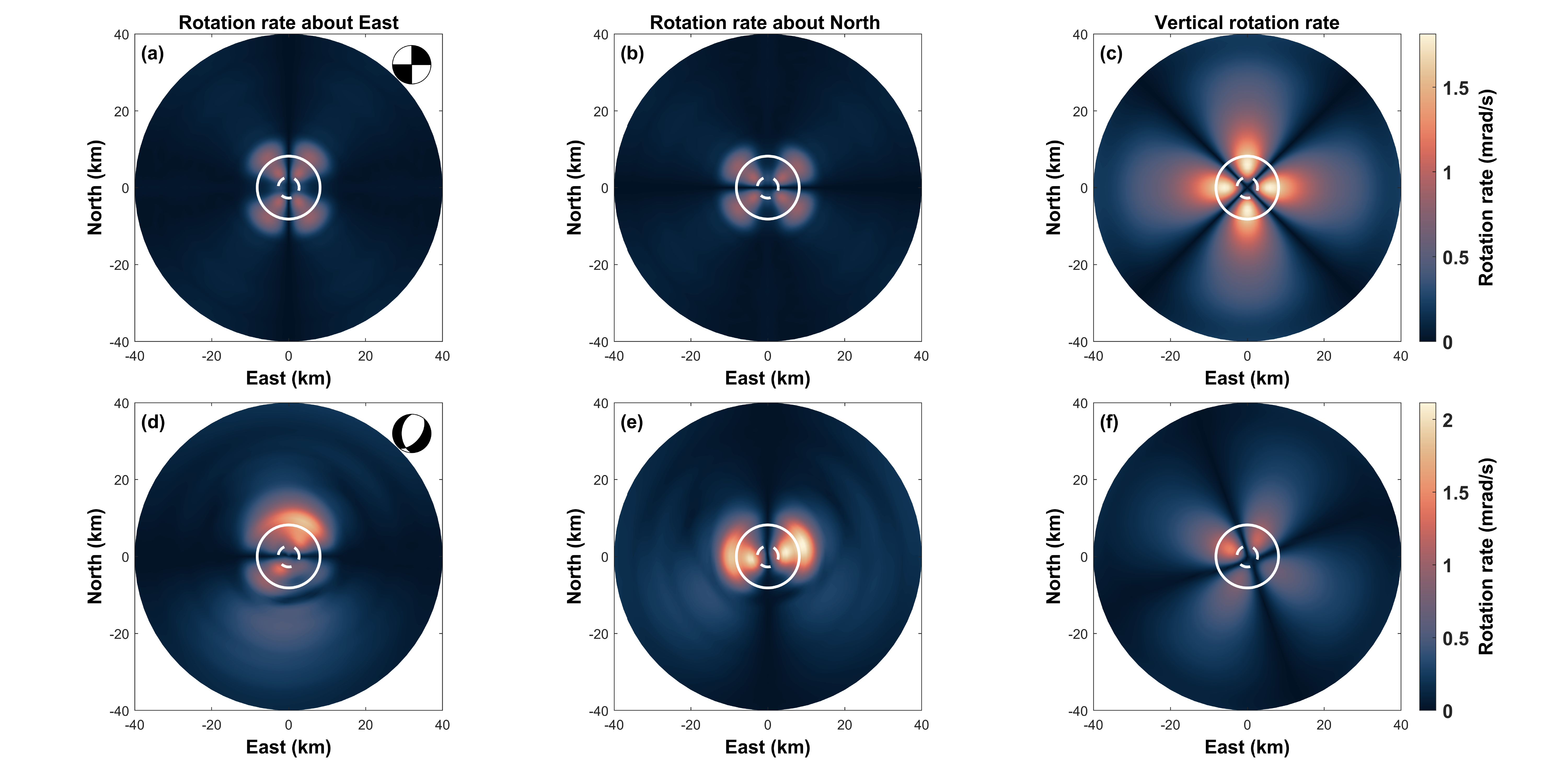}
	\caption{Comparison of PGRR maps about the three orthogonal directions for  layered medium subjected to two different source mechanisms \textbf{(a--c)} Strike-slip fault and \textbf{(d--f)} Oblique normal fault . The dotted white line marks the end of near field zone while the solid white line marks the end of intermediate field zone.}
	\label{comparison_PGRR_all_sources}
\end{figure*}

To examine the influence of source mechanism in the layered medium, six representative double-couple earthquake sources with same moment magnitude are considered. For brevity, the discussion presented here focuses on the comparison between a left-lateral strike-slip source (strike = \SI{0}{\degree}, dip = \SI{90}{\degree}, rake = \SI{0}{\degree}) and an oblique-normal source (strike = \SI{0}{\degree}, dip = \SI{45}{\degree}, rake = \SI{-120}{\degree}). Figure~\ref{comparison_PGRR_all_sources} compares the spatial distributions of PGRR about the three orthogonal axes for these two source mechanisms and describes the evolution of rotational demand within the near-, intermediate-, and far-wavefield regions.

The propagation patterns of PGRR for the left-lateral strike-slip source (Figure \ref{comparison_PGRR_all_sources}a--\ref{comparison_PGRR_all_sources}c) exhibit the same fundamental characteristics as observed previously for the homogeneous medium (Figure \ref{PGV_vs_PGRR}a--\ref{PGV_vs_PGRR}c). The tilt components retain the quadruple radiation patterns with zero rotation along nodal and auxiliary plane. The torsional component continues to exhibit a symmetric quadruple-lobed distribution aligned with the fault geometry. This demonstrates that the primary features of rotational wave propagation are governed by the earthquake source mechanism and remain largely unaffected by the introduction of a generic layered medium. The layered velocity structure, however, produces a noticeable increase in the PGRR amplitudes together with a slight distortion of the lobe geometry. These modifications arise from wave reflections, refractions, and mode conversions at the layer interfaces. The interfaces locally enhance the spatial gradients of the displacement field responsible for rotational ground motions.

The oblique-normal fault (Figures \ref{comparison_PGRR_all_sources}d--\ref{comparison_PGRR_all_sources}f), as expected, produces markedly different propagation characteristics. The tilt components exhibit predominantly double-lobed radiation patterns with unequal lobe amplitudes. This feature reflects the asymmetric radiation generated by the combined strike-slip and dip-slip mechanism of the source. Unlike the strike-slip source, the orientation of the dominant lobes cannot be directly associated with the nodal or auxiliary planes, indicating that the additional dip-slip component substantially alters the rotational radiation pattern. The torsional component also exhibits a quadruple distribution; however, both the orientation and the relative amplitudes of the four lobes differ significantly from those of the strike-slip fault. The strongest rotational amplitudes are aligned approximately along the rake angle, indicating that the torsional component retains information about the dominant slip direction. These observations demonstrate that rotational ground motions are highly sensitive to earthquake source mechanism and preserve detailed characteristics of the seismic radiation pattern that may be exploited for improved source characterization.

To further investigate the spatial evolution of rotational ground motions, the near-, intermediate-, and far-wavefield regions are indicated in Figure (\ref{comparison_PGRR_all_sources}). The inner dashed white contour denotes the approximate extent of the near field, corresponding to one dominant wavelength of the source, whereas the outer solid white contour marks the boundary of the intermediate field at approximately three dominant wavelengths. Beyond this distance, the wavefield is considered to be in the far-field regime. Distinct differences are observed in the propagation of the rotational components for the two source mechanisms. For the strike-slip fault, the torsional component propagates well beyond the intermediate-wavefield region and remains significant into the far field, whereas the tilt components are largely confined to the near- and intermediate-wavefield regions. In contrast, the oblique-normal fault exhibits the opposite behavior. The tilt components retain substantial amplitudes beyond the intermediate-wavefield boundary and the torsional component decays more rapidly with distance. These results demonstrate that the earthquake source mechanism governs not only the amplitude and spatial distribution of PGRR but also the dominant type of rotational motion that persists during propagation. Consequently, different faulting mechanisms may generate markedly different combinations of tilt and torsional demand, suggesting that both the amplitude and the nature of rotational ground motions should be considered in future assessments of rotational seismic hazard and the seismic design of structures.
	
\section{Rotational Ground Motions in the Indo-Ganga Basin}
The analyses presented in the previous section demonstrates that the amplitude and propagation patterns of rotational ground motions are strongly influenced by the earthquake source mechanism. While these observations provide fundamental insights into rotational wave propagation in idealized layered media, actual earthquake ground motions are further modified by complex geological structures like sedimentary basins. Owing to the dependence of rotational motions on spatial displacement gradients, they are expected to be sensitive to the basin effects. The present section therefore investigates the propagation of rotational ground motions in the Indo-Gangetic basin.

The Indo-Gangetic basin, located along the southern foothills of the Himalayas, is one of the largest sedimentary basins in the world and lies within one of the most seismically active regions of the Indian subcontinent. The basin has experienced several damaging earthquakes and supports a high population density, making it vulnerable to seismic hazards \citep{Dhabu2021b}. Previous studies have extensively investigated the influence of the deep sedimentary deposits and the adjoining Himalayan region on translational ground motions \citep{Jayalakshmi2020, Dhabu2021b}. However, the characteristics of rotational ground motions in this basin region have remained largely unexplored. 
	
To investigate the effect of sedimentary basin on rotational ground motions, a 3D spectral-element model of the basin is developed using the material properties listed in Table \ref{IG_properties}.
\begin{table}
	\tbl{Material properties of the Indo-Ganga basin \citep{Jayalakshmi2020}\label{IG_properties}}%
	{\begin{tabular*}{\columnwidth}{@{\extracolsep{\fill}}ccccc@{}} %
			Layer & Depth & $V_s$ & $V_p$ & Density    \\
			& (km)  & (m/s) & (m/s) & (kg/m$^3$) \\ \hline
			% \multicolumn{1}{l}{} & \textbf{12856} & \textbf{23.2 \%} & \textbf{0.2 \%} & \textbf{18.8 \%} \\
			\colrule
			1    & 0 -- 0.4     & 475   &  740  & 1568       \\
			2    & 0.4 -- 2.0   & 900   & 1400  & 1897       \\
			3    & 2.0 -- 3.6   & 2100  & 3300  & 2335       \\
			4    & 3.6 -- 5.2   & 2700  & 4200  & 2494       \\
			5    & 5.2 -- 6.8   & 3040  & 4880  & 2587       \\
			6    & 6.8 -- 30.0  & 3200  & 5600  & 2680       \\
			\botrule
	\end{tabular*}}
	{}
\end{table}
It should be noted that the present model does not consider the effect of topography. Owing to the very low near-surface shear-wave velocities within the basin sediments, a highly refined numerical discretization is required to accurately resolve seismic-wave propagation up to the target frequency range. The resulting computational model consists of approximately $26.2$ million spectral elements, with a maximum element size of \SI{200}{\meter}, corresponding to a minimum resolvable period of \SI{0.52}{\second}. The developed model is used to simulate rotational ground motions for the same source-receiver configuration and the two representative source mechanisms (strike-slip and oblique-normal source) as considered in the previous section. Each simulation requires approximately 6 hours of wall-clock time on 512 cores. The simulated ground motions are subsequently band-pass filtered between \SI{0.25}{\hertz} and \SI{1.9}{\hertz}. 

\begin{figure*} % fig1
	\global\twocolcaptionfalse
	\centering
	\includegraphics[width=\textwidth, keepaspectratio]{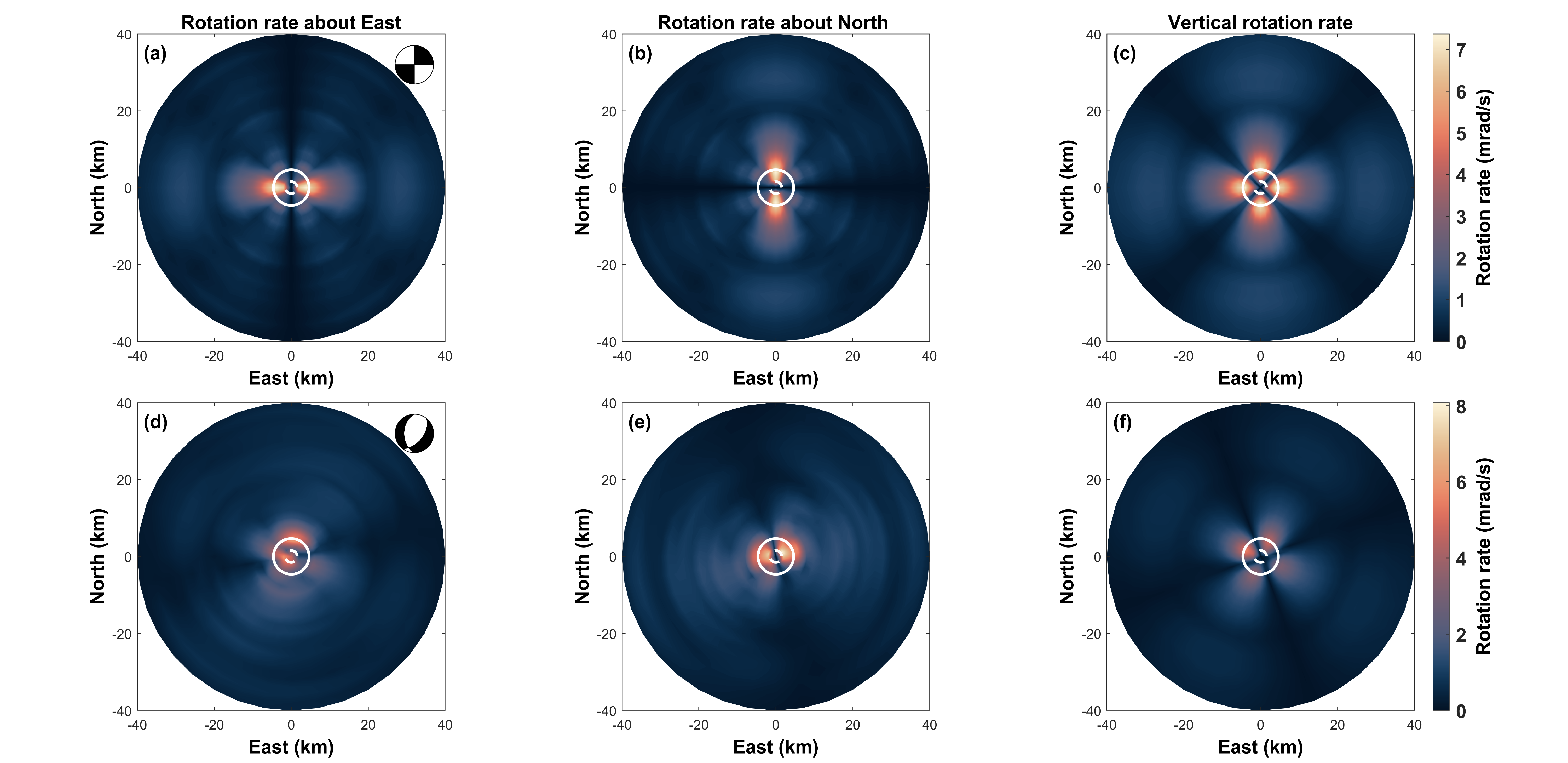}
	\caption{Comparison of PGRR maps about the three orthogonal directions for the two source mechanisms obtained for the Indo-Ganga basin \textbf{(a--c)} Strike-slip fault and \textbf{(d--f)} Oblique normal fault. }
	\label{IG_basin_PGRR}
\end{figure*}

Figure (\ref{IG_basin_PGRR}) presents the resulting PGRR maps for the two considered earthquake source mechanisms.
In comparison with the generic layered medium, the basin produces a pronounced amplification of PGRR together with significant changes in the spatial distribution of rotational demand. These modifications arise from the low P-wave velocity and S-wave velocity layers and also stronger inter-layer contrast. They together alter the spatial gradients of the seismic wavefield responsible for generating rotations.

For the strike-slip source (Figures \ref{IG_basin_PGRR}a--\ref{IG_basin_PGRR}c), the overall characteristics identified for the generic layered medium remain recognizable, but the amplitudes increase considerably. The maximum PGRR reaches approximately \SI{7.1}{mrad/s}, compared with only \SI{1.84}{mrad/s} for the generic layered medium, corresponding to an amplification of nearly four times. The quadruple-lobed radiation patterns observed for the tilt components in the generic layered medium (Figure \ref{comparison_PGRR_all_sources}a--\ref{comparison_PGRR_all_sources}b) evolve into broader double-lobed patterns. The dominant lobes are now aligned along the nodal planes in contrast to zero rotations along the nodal planes in generic layered medium. It should be noted that the physical interpretation developed for the formation of dominant lobes in the homogeneous model (Figure \ref{abs_signed_comparison}) is no longer sufficient to explain the observed propagation patterns because the rotational wavefield is now governed by the combined effects of source radiation and basin-induced wave propagation. The torsional component retains an overall quadruple radiation pattern but develops a pronounced region of reduced rotational demand in the central far-field, indicating destructive interference caused by wave scattering and multiple reflections within the basin sediments.

The influence of the basin is even more pronounced for the oblique-normal fault (Figures \ref{IG_basin_PGRR}d--\ref{IG_basin_PGRR}f). The maximum PGRR increases from approximately \SI{2.1}{mrad/s} in the generic layered medium to about \SI{8.0}{mrad/s} in the basin model. In addition to amplification, the propagation characteristics of the tilt components are fundamentally altered. The double-lobed radiation pattern observed in the generic layered medium (Figure \ref{comparison_PGRR_all_sources}d--\ref{comparison_PGRR_all_sources}e) evolves into a weak quadruple pattern for rotation about the East axis. The rotation about the North-axis retains the double-lobed distribution but with a modified orientation of the zero-rotation direction that approximately follows the rake direction of the fault. In contrast, the torsional component shows comparatively minor changes in its overall radiation pattern, suggesting that basin exhibits stronger effects on the tilt components as compared to torsion for oblique normal source mechanism.

For both the source mechanisms, significant rotational amplitudes persist well beyond the intermediate-wavefield region, demonstrating that basin-induced amplification is not confined to the near-source region but influences rotational motions over large propagation distances. The dominant rotational demand differs between the two source mechanisms. For the oblique-normal fault, the rotational response is governed primarily by the tilt components, whereas for the strike-slip source both tilt and torsional motions contribute significantly to the overall rotational demand. These results demonstrate that realistic sedimentary basins modify not only the amplitude but also the dominant type of rotational motion experienced at the ground surface. Simplified homogeneous or layered Earth models are thus likely to underestimate both the complexity and the magnitude of rotational ground motions, emphasizing the importance of incorporating realistic basin structures in rotational seismic hazard assessment.

\section{Conclusions}
The study presents a numerical framework to simulate and analyze rotational ground motions generated by earthquakes in 3D geological media. The open-source spectral-element package SPECFEM3D was extended to compute rotational waveforms directly within the numerical solver. This implementation eliminates the dependence of determining rotations on receiver spacing and post-processing procedures, and enables physically consistent computation of rotational wavefields. The conducted validation against analytical solutions for a homogeneous elastic medium demonstrated excellent agreement in waveform characteristics, amplitude, and spatial propagation that confirms the accuracy and robustness of the implementation.

The developed framework was subsequently employed to investigate the physical characteristics of rotational ground motions. Comparison of translational and rotational waveforms demonstrated that rotational components possess distinct wavefield sensitivities. The present work introduces the concept of Peak Ground Rotation Rate (PGRR) maps to characterize the spatial distribution of maximum rotational demand. The PGRR maps revealed that the propagation characteristics of rotational motions depend strongly on earthquake source mechanism. As rotations are governed by the spatial gradients of the seismic wavefield rather than the wavefield itself, they show propagation patterns that contrast with those of translational motion. Consequently, the regions that experience maximum rotations do not coincide with the regions that undergo maximum translational motions. The influence of geological complexity was further investigated using a generic layered Earth model and a realistic 3D model of the Indo-Gangetic basin. While the layered medium preserved the fundamental propagation characteristics observed in the homogeneous model, realistic basin sediments substantially modified both the amplitude and spatial distribution of rotational ground motions. Comparisons between strike-slip and oblique-normal source mechanisms demonstrated that both the amplitude and the dominant rotational component vary significantly with focal mechanism. This observation emphasized the importance of accounting the rotational ground motions in future hazard assessment and earthquake-resistant design of structures.

Thus, the present work provides a foundation for future studies on rotational seismic hazard, earthquake source characterization, soil-structure interaction, and the development of earthquake-resistant structural design methodologies to sustain rotations. As rotational sensing technology continues to advance, the numerical framework presented here offers a valuable tool for identifying locations of high rotational demand. As dense rotational seismic networks become increasingly available, this framework will also aid in interpreting field observations and in understanding the role of rotational motions in future earthquakes.

	\begin{datres}%[CUSTOM HEAD]
		The extended version of SPECFEM3D developed in this study is currently available as a private Git repository and can be obtained by contacting the first author. The authors contacted the SPECFEM3D development team to verify whether the present rotational-motion implementation is included in the current official version of the software and were informed that it is not (SPECFEM3D Development Team, personal communication). Following acceptance of this work, the authors and the SPECFEM3D development team intend to make the extension publicly available through the official SPECFEM3D repository.
	\end{datres}

	\section{Declaration of Competing Interests}
	
	The authors acknowledge that there are no conflicts of interest recorded.\footnote{The authors acknowledge that there are no conflicts of interest recorded.}
	
	\begin{ack}%[CUSTOM ACK HEAD]
		This work was funded by the Deutsche Forschungsgemeinschaft (DFG, German Research Foundation) – Project number 521891289.
		
		This work used resources of the Deutsches Klimarechenzentrum (DKRZ) granted by its Scientific Steering Committee (WLA) under project ID 1326.
	\end{ack}
	
	\bibliography{HERS_paper1}

@Article{Dhabu2021,
  author    = {Dhabu, Anjali C. and Singla, Varun K. and STG, Raghukanth},
  journal   = {Journal of Geophysical Research: Solid Earth},
  title     = {Seismic Wave Propagation Through Layered Reduced Micropolar Medium},
  year      = {2021},
  issn      = {2169-9356},
  month     = Nov,
  number    = {11},
  volume    = {126},
  doi       = {10.1029/2020jb020931},
  publisher = {American Geophysical Union (AGU)},
}

@Article{Cao2024,
  author    = {Cao, Yenan and Mavroeidis, George P.},
  journal   = {Journal of Seismology},
  title     = {Numerical simulation of near-fault ground strains and rotations from actual earthquakes with predominantly dip-slip mechanisms},
  year      = {2024},
  issn      = {1573-157X},
  month     = July,
  number    = {4},
  pages     = {1027--1053},
  volume    = {28},
  doi       = {10.1007/s10950-024-10230-7},
  publisher = {Springer Science and Business Media LLC},
}

@Article{Cao2019,
  author    = {Cao, Yenan and Mavroeidis, George P.},
  journal   = {Bulletin of the Seismological Society of America},
  title     = {A Parametric Investigation of Near‐Fault Ground Strains and Rotations Using Finite‐Fault Simulations},
  year      = {2019},
  issn      = {1943-3573},
  month     = Sept,
  number    = {5},
  pages     = {1758--1784},
  volume    = {109},
  doi       = {10.1785/0120190045},
  publisher = {Seismological Society of America (SSA)},
}

@Article{Komatitsch2002,
  author    = {Komatitsch, Dimitri and Tromp, Jeroen},
  journal   = {Geophysical Journal International},
  title     = {Spectral-element simulations of global seismic wave propagation-I. Validation},
  year      = {2002},
  issn      = {1365-246X},
  month     = May,
  number    = {2},
  pages     = {390--412},
  volume    = {149},
  doi       = {10.1046/j.1365-246x.2002.01653.x},
  publisher = {Oxford University Press (OUP)},
}

@Article{Komatitsch2002a,
  author    = {Komatitsch, Dimitri and Tromp, Jeroen},
  journal   = {Geophysical Journal International},
  title     = {Spectral-element simulations of global seismic wave propagation-II. Three-dimensional models, oceans, rotation and self-gravitation},
  year      = {2002},
  issn      = {1365-246X},
  month     = July,
  number    = {1},
  pages     = {303--318},
  volume    = {150},
  doi       = {10.1046/j.1365-246x.2002.01716.x},
  publisher = {Oxford University Press (OUP)},
}

@Misc{Komatitsch2024,
  author    = {Komatitsch, Dimitri and Tromp, Jeroen and Garg, Rahul and Gharti, Hom Nath and Nagaso, Masaru and Oral, Elif and Peter, Daniel and Afanasiev, Michael and Almada, Rafael and Ampuero, Jean-Paul and Bachmann, Etienne and Bai, Kangchen and Basini, Piero and Beller, Stephen and Bishop, Jordan and Bissey, Francois and Blitz, Celine and Bottero, Alexis and Bozdag, Ebru and Casarotti, Emanuele and Charles, Joseph and Chen, Min and Cristini, Paul and Durochat, Clement and Galvez Barron, Percy and Gassmoeller, Rene and Goeddeke, Dominik and Grinberg, Leopold and Gupta, Aakash and Heien, Eric and Hjoerleifsdottir, Vala and Karakostas, Foivos and Kientz, Sue and Labarta, Jesus and Le Goff, Nicolas and Le Loher, Pieyre and Lefebvre, Matthieu and Liu, Qinya and Liu, Youshan and Luet, David and Luo, Yang and Maggi, Alessia and Magnoni, Federica and Martin, Roland and Matzen, Rene and McBain, G. D. and McRitchie, Dennis and Meschede, Matthias and Messmer, Peter and Michea, David and Miller, David and Modrak, Ryan and Monteiller, Vadim and Morency, Christina and Nadh Somala, Surendra and Nissen-Meyer, Tarje and Pouget, Kevin and Rietmann, Max and Sales de Andrade, Elliott and Savage, Brian and Schuberth, Bernhard and Sieminski, Anne and Smith, James and Strand, Leif and Tape, Carl and Valero Cano, Eduardo and Videau, Brice and Vilotte, Jean-Pierre and Weng, Huihui and Xie, Zhinan and Zhang, Chang-Hua and Zhu, Hejun},
  title     = {SPECFEM/specfem3d: SPECFEM3D v4.1.1},
  year      = {2024},
  copyright = {GNU General Public License v3.0 only},
  doi       = {10.5281/ZENODO.10823181},
  publisher = {Zenodo},
}

@Article{Dhabu2021a,
  author    = {Dhabu, Anjali C. and Raghukanth, S. T. G.},
  journal   = {Journal of Seismology},
  title     = {Translational and rotational ground motion simulations in homogeneous reduced micropolar half-space},
  year      = {2021},
  issn      = {1573-157X},
  month     = Mar,
  number    = {2},
  pages     = {599--623},
  volume    = {25},
  doi       = {10.1007/s10950-021-09983-2},
  publisher = {Springer Science and Business Media LLC},
}

@Article{Singla2019,
  author    = {Singla, Varun K. and Gupta, Vinay K.},
  journal   = {Bulletin of the Seismological Society of America},
  title     = {Surface Rotations Due to Kinematic Shear Dislocation Point Source in a Multilayered Elastic Medium},
  year      = {2019},
  issn      = {1943-3573},
  month     = Jan,
  number    = {1},
  pages     = {433--447},
  volume    = {109},
  doi       = {10.1785/0120180083},
  publisher = {Seismological Society of America (SSA)},
}

@Article{Dhabu2021b,
  author    = {Dhabu, Anjali Chandrashekhar and Gudimella, Raghukanth Srimath Tirumala},
  journal   = {Arabian Journal of Geosciences},
  title     = {Influence of Himalayan topography on earthquake ground motions},
  year      = {2021},
  issn      = {1866-7538},
  month     = Sept,
  number    = {18},
  volume    = {14},
  doi       = {10.1007/s12517-021-08111-1},
  publisher = {Springer Science and Business Media LLC},
}

@Article{Jayalakshmi2020,
  author    = {Jayalakshmi, S. and Dhanya, J. and Raghukanth, S.T.G. and Martin Mai, P.},
  journal   = {Soil Dynamics and Earthquake Engineering},
  title     = {3D seismic wave amplification in the Indo-Gangetic basin from spectral element simulations},
  year      = {2020},
  issn      = {0267-7261},
  month     = Feb,
  pages     = {105923},
  volume    = {129},
  doi       = {10.1016/j.soildyn.2019.105923},
  publisher = {Elsevier BV},
}

@Article{Suryanto2006,
  author    = {Suryanto, W. and Igel, H. and Wassermann, J. and Cochard, A. and Schuberth, B. and Vollmer, D. and Scherbaum, F. and Schreiber, U. and Velikoseltsev, A.},
  journal   = {Bulletin of the Seismological Society of America},
  title     = {First Comparison of Array-Derived Rotational Ground Motions with Direct Ring Laser Measurements},
  year      = {2006},
  issn      = {0037-1106},
  month     = Dec,
  number    = {6},
  pages     = {2059--2071},
  volume    = {96},
  doi       = {10.1785/0120060004},
  publisher = {Seismological Society of America (SSA)},
}

@Article{Lin2012,
  author    = {Lin, Chin-Jen and Huang, Wen-Gee and Huang, Han-Pang and Huang, Bor-Shouh and Ku, Chin-Shang and Liu, Chun-Chi},
  journal   = {Journal of Seismology},
  title     = {Investigation of array-derived rotation in TAIPEI 101},
  year      = {2012},
  issn      = {1573-157X},
  month     = May,
  number    = {4},
  pages     = {721--731},
  volume    = {16},
  doi       = {10.1007/s10950-012-9306-7},
  publisher = {Springer Science and Business Media LLC},
}

@Article{Taylor2021,
  author    = {Taylor, G. and Hillers, G. and Vuorinen, T. A. T.},
  journal   = {Geophysical Research Letters},
  title     = {Using Array‐Derived Rotational Motion to Obtain Local Wave Propagation Properties From Earthquakes Induced by the 2018 Geothermal Stimulation in Finland},
  year      = {2021},
  issn      = {1944-8007},
  month     = Mar,
  number    = {6},
  volume    = {48},
  doi       = {10.1029/2020gl090403},
  publisher = {American Geophysical Union (AGU)},
}

@Article{Stedman1995,
  author    = {Stedman, G. E. and Li, Z. and Bilger, H. R.},
  journal   = {Applied Optics},
  title     = {Sideband analysis and seismic detection in a large ring laser},
  year      = {1995},
  issn      = {1539-4522},
  month     = Aug,
  number    = {24},
  pages     = {5375},
  volume    = {34},
  doi       = {10.1364/ao.34.005375},
  publisher = {Optica Publishing Group},
}

@Article{Pancha2000,
  author    = {Pancha, A. and Webb, T. H. and Stedman, G. E. and McLeod, D. P. and Schreiber, K. U.},
  journal   = {Geophysical Research Letters},
  title     = {Ring laser detection of rotations from teleseismic waves},
  year      = {2000},
  issn      = {1944-8007},
  month     = Nov,
  number    = {21},
  pages     = {3553--3556},
  volume    = {27},
  doi       = {10.1029/2000gl011734},
  publisher = {American Geophysical Union (AGU)},
}

@InCollection{schreiber2006geosensor,
  author    = {Schreiber, Ulrich and Igel, Heiner and Cochard, Alain and Velikoseltsev, Alexander and Flaws, Asher and Schuberth, Bernhard and Drewitz, Wolfgang and M{\"u}ller, Frieder},
  booktitle = {Observation of the Earth System from Space},
  publisher = {Springer},
  title     = {The GEOsensor project: rotations—a new observable for seismology},
  year      = {2006},
  pages     = {427--443},
}

@Book{aki2002quantitative,
  author    = {Aki, Keiiti and Richards, Paul},
  publisher = {MIT Press},
  title     = {Quantitative seismology},
  year      = {2002},
}

@Book{chopra2007dynamics,
  author    = {Chopra, Anil K},
  publisher = {Pearson Education India},
  title     = {Dynamics of structures},
  year      = {2007},
}

@Article{Bormann2012,
  author    = {Bormann, Peter and Engdahl, E. Robert and Kind, Rainer},
  journal   = {New Manual of Seismological Observatory Practice 2 (NMSOP2)},
  title     = {Seismic Wave Propagation and Earth models},
  year      = {2012},
  doi       = {10.2312/GFZ.NMSOP-2_CH2},
  language  = {en},
  publisher = {Deutsches GeoForschungsZentrum GFZ},
}

@Article{beroza1995seismic,
  author    = {Beroza, Gregory C},
  journal   = {Reviews of Geophysics},
  title     = {Seismic source modeling},
  year      = {1995},
  number    = {S1},
  pages     = {299--308},
  volume    = {33},
  publisher = {Wiley Online Library},
}

@Article{Tromp2019,
  author    = {Tromp, Jeroen},
  journal   = {Nature Reviews Earth \&amp; Environment},
  title     = {Seismic wavefield imaging of Earth’s interior across scales},
  year      = {2019},
  issn      = {2662-138X},
  month     = Dec,
  number    = {1},
  pages     = {40--53},
  volume    = {1},
  doi       = {10.1038/s43017-019-0003-8},
  publisher = {Springer Science and Business Media LLC},
}

@Article{Tkalcic2022,
  author    = {Tkalčić, Hrvoje and Wang, Sheng and Ph\d{a}m, Thanh-Son},
  journal   = {Annual Review of Earth and Planetary Sciences},
  title     = {Shear Properties of Earth’s Inner Core},
  year      = {2022},
  issn      = {1545-4495},
  month     = May,
  number    = {1},
  pages     = {153--181},
  volume    = {50},
  doi       = {10.1146/annurev-earth-071521-063942},
  publisher = {Annual Reviews},
}

@Article{Gruenthal2018,
  author    = {Grünthal, Gottfried and Stromeyer, Dietrich and Bosse, Christian and Cotton, Fabrice and Bindi, Dino},
  journal   = {Bulletin of Earthquake Engineering},
  title     = {The probabilistic seismic hazard assessment of Germany—version 2016, considering the range of epistemic uncertainties and aleatory variability},
  year      = {2018},
  issn      = {1573-1456},
  month     = Feb,
  number    = {10},
  pages     = {4339--4395},
  volume    = {16},
  doi       = {10.1007/s10518-018-0315-y},
  publisher = {Springer Science and Business Media LLC},
}

@Article{Nath2012,
  author    = {Nath, S. K. and Thingbaijam, K. K. S.},
  journal   = {Seismological Research Letters},
  title     = {Probabilistic Seismic Hazard Assessment of India},
  year      = {2012},
  issn      = {1938-2057},
  month     = Jan,
  number    = {1},
  pages     = {135--149},
  volume    = {83},
  doi       = {10.1785/gssrl.83.1.135},
  publisher = {Seismological Society of America (SSA)},
}

@Article{Lee2009,
  author    = {Lee, W. H. K. and Igel, H. and Trifunac, M. D.},
  journal   = {Seismological Research Letters},
  title     = {Recent Advances in Rotational Seismology},
  year      = {2009},
  issn      = {0895-0695},
  month     = May,
  number    = {3},
  pages     = {479--490},
  volume    = {80},
  doi       = {10.1785/gssrl.80.3.479},
  publisher = {Seismological Society of America (SSA)},
}

@Article{Bernauer2012,
  author    = {Bernauer, Felix and Wassermann, Joachim and Igel, Heiner},
  journal   = {Journal of Seismology},
  title     = {Rotational sensors—a comparison of different sensor types},
  year      = {2012},
  issn      = {1573-157X},
  month     = Mar,
  number    = {4},
  pages     = {595--602},
  volume    = {16},
  doi       = {10.1007/s10950-012-9286-7},
  publisher = {Springer Science and Business Media LLC},
}

@Article{Bernauer2018,
  author    = {Bernauer, Felix and Wassermann, Joachim and Guattari, Frederic and Frenois, Arnaud and Bigueur, Alexandre and Gaillot, Arnaud and de Toldi, Elliot and Ponceau, Damien and Schreiber, Ulrich and Igel, Heiner},
  journal   = {Seismological Research Letters},
  title     = {BlueSeis3A: Full Characterization of a 3C Broadband Rotational Seismometer},
  year      = {2018},
  issn      = {1938-2057},
  month     = Jan,
  number    = {2A},
  pages     = {620--629},
  volume    = {89},
  doi       = {10.1785/0220170143},
  publisher = {Seismological Society of America (SSA)},
}

@Article{Salvermoser2017,
  author    = {Salvermoser, Johannes and Hadziioannou, Céline and Hable, Sarah and Krischer, Lion and Chow, Bryant and Ramos, Catalina and Wassermann, Joachim and Schreiber, Ulrich and Gebauer, André and Igel, Heiner},
  journal   = {Seismological Research Letters},
  title     = {An Event Database for Rotational Seismology},
  year      = {2017},
  issn      = {1938-2057},
  month     = Mar,
  number    = {3},
  pages     = {935--941},
  volume    = {88},
  doi       = {10.1785/0220160184},
  publisher = {Seismological Society of America (SSA)},
}

@Article{bouchon1982strain,
  author    = {Bouchon, Michel and Aki, Keiiti},
  journal   = {Bulletin of the Seismological Society of America},
  title     = {Strain, tilt, and rotation associated with strong ground motion in the vicinity of earthquake faults},
  year      = {1982},
  number    = {5},
  pages     = {1717--1738},
  volume    = {72},
  publisher = {The Seismological Society of America},
}

@Article{Takeo1998,
  author    = {Takeo, Minoru},
  journal   = {Geophysical Research Letters},
  title     = {Ground rotational motions recorded in near‐source region of earthquakes},
  year      = {1998},
  issn      = {1944-8007},
  month     = Mar,
  number    = {6},
  pages     = {789--792},
  volume    = {25},
  doi       = {10.1029/98gl00511},
  publisher = {American Geophysical Union (AGU)},
}

@Article{nigbor1994six,
  author    = {Nigbor, Robert L},
  journal   = {Bulletin of the Seismological Society of America},
  title     = {Six-degree-of-freedom ground-motion measurement},
  year      = {1994},
  number    = {5},
  pages     = {1665--1669},
  volume    = {84},
  publisher = {The Seismological Society of America},
}

@Article{Sollberger2020,
  author    = {Sollberger, David and Igel, Heiner and Schmelzbach, Cedric and Edme, Pascal and van Manen, Dirk-Jan and Bernauer, Felix and Yuan, Shihao and Wassermann, Joachim and Schreiber, Ulrich and Robertsson, Johan O. A.},
  journal   = {Sensors},
  title     = {Seismological Processing of Six Degree-of-Freedom Ground-Motion Data},
  year      = {2020},
  issn      = {1424-8220},
  month     = Dec,
  number    = {23},
  pages     = {6904},
  volume    = {20},
  doi       = {10.3390/s20236904},
  publisher = {MDPI AG},
}

@Article{Chen2023,
  author    = {Chen, Chang and Wang, Yun and Sun, Lixia and Lin, Chin‐Jen and Wei, Yongxiang and Liao, Chunqi and Lin, Binhua and Qin, Linpeng},
  journal   = {Earth and Space Science},
  title     = {Six‐Component Earthquake Synchronous Observations Across Taiwan Strait: Phase Velocity and Source Location},
  year      = {2023},
  issn      = {2333-5084},
  month     = Dec,
  number    = {12},
  volume    = {10},
  doi       = {10.1029/2023ea003040},
  publisher = {American Geophysical Union (AGU)},
}

@Article{Xu2025,
  author    = {Xu, Beibei and Zhang, Yong and Li, Wei and Chen, Chang and Wang, Yun},
  journal   = {Seismological Research Letters},
  title     = {Incorporating Directly Observed Rotational Seismic Data into Moment Tensor Inversion for the 2019 Mw 6.2 Hualien Earthquake},
  year      = {2025},
  issn      = {1938-2057},
  month     = July,
  number    = {1},
  pages     = {15--27},
  volume    = {97},
  doi       = {10.1785/0220240470},
  publisher = {Seismological Society of America (SSA)},
}

@InProceedings{newmark1969torsion,
  author       = {Newmark, Nathan M and others},
  title        = {Torsion in symmetrical buildings},
  year         = {1969},
  organization = {PROCEEDING OF WORLD CONFERENCE ON EARTHQUAKE ENGINEERING},
}

@Article{Chandler1986,
  author    = {Chandler, Adrian M.},
  journal   = {Nature},
  title     = {Building damage in Mexico City earthquake},
  year      = {1986},
  issn      = {1476-4687},
  month     = Apr,
  number    = {6062},
  pages     = {497--501},
  volume    = {320},
  doi       = {10.1038/320497a0},
  publisher = {Springer Science and Business Media LLC},
}

@Article{Gueguen2020,
  author    = {Guéguen, Philippe and Guattari, Frédéric and Aubert, Coralie and Laudat, Theo},
  journal   = {Sensors},
  title     = {Comparing Direct Observation of Torsion with Array-Derived Rotation in Civil Engineering Structures},
  year      = {2020},
  issn      = {1424-8220},
  month     = Dec,
  number    = {1},
  pages     = {142},
  volume    = {21},
  doi       = {10.3390/s21010142},
  publisher = {MDPI AG},
}

@Article{Smerzini2009,
  author    = {Smerzini, Chiara and Paolucci, Roberto and Stupazzini, Marco},
  journal   = {Journal of Earthquake Engineering},
  title     = {Experimental and Numerical Results on Earthquake-Induced Rotational Ground Motions},
  year      = {2009},
  issn      = {1559-808X},
  month     = Apr,
  number    = {sup1},
  pages     = {66--82},
  volume    = {13},
  doi       = {10.1080/13632460902813299},
  publisher = {Informa UK Limited},
}

@Article{Agnew2014,
  author    = {Agnew, D. C. and Wyatt, F. K.},
  journal   = {Bulletin of the Seismological Society of America},
  title     = {Dynamic Strains at Regional and Teleseismic Distances},
  year      = {2014},
  issn      = {0037-1106},
  month     = July,
  number    = {4},
  pages     = {1846--1859},
  volume    = {104},
  doi       = {10.1785/0120140007},
  publisher = {Seismological Society of America (SSA)},
}

@Article{Schmelzbach2018,
  author    = {Schmelzbach, Cedric and Donner, Stefanie and Igel, Heiner and Sollberger, David and Taufiqurrahman, Taufiq and Bernauer, Felix and Häusler, Mauro and Van Renterghem, Cédéric and Wassermann, Joachim and Robertsson, Johan},
  journal   = {Geophysics},
  title     = {Advances in 6C seismology: Applications of combined translational and rotational motion measurements in global and exploration seismology},
  year      = {2018},
  issn      = {1942-2156},
  month     = Apr,
  number    = {3},
  pages     = {WC53--WC69},
  volume    = {83},
  doi       = {10.1190/geo2017-0492.1},
  publisher = {Society of Exploration Geophysicists},
}

@Article{Gade2015,
  author    = {Gade, Maheshreddy and Raghukanth, S.T.G.},
  journal   = {Applied Mathematical Modelling},
  title     = {Seismic ground motion in micropolar elastic half-space},
  year      = {2015},
  issn      = {0307-904X},
  month     = Dec,
  number    = {23-24},
  pages     = {7244--7265},
  volume    = {39},
  doi       = {10.1016/j.apm.2015.03.006},
  publisher = {Elsevier BV},
}

@Article{Dhabu2019,
  author    = {Dhabu, Anjali C. and Raghukanth, S. T. G.},
  journal   = {Pure and Applied Geophysics},
  title     = {Fundamental Solutions to Static and Dynamic Loads for Homogeneous Reduced Micropolar Half-Space},
  year      = {2019},
  issn      = {1420-9136},
  month     = June,
  number    = {11},
  pages     = {4881--4905},
  volume    = {176},
  doi       = {10.1007/s00024-019-02225-0},
  publisher = {Springer Science and Business Media LLC},
}
	
\end{document}